\documentclass[english,onecolumn]{IEEEtran}
\usepackage[T1]{fontenc}
\usepackage[latin9]{inputenc}
\usepackage{amsmath}
\usepackage{amsthm}
\usepackage{amssymb}
\usepackage{graphicx}
\usepackage{esint}

\makeatletter
\theoremstyle{plain}
\newtheorem{thm}{\protect\theoremname}
\theoremstyle{definition}
\newtheorem{defn}[thm]{\protect\definitionname}
\theoremstyle{plain}
\newtheorem{prop}[thm]{\protect\propositionname}
\theoremstyle{plain}
\newtheorem{cor}[thm]{\protect\corollaryname}
\theoremstyle{plain}
\newtheorem{conjecture}[thm]{\protect\conjecturename}

\usepackage{mathrsfs}
\usepackage{algorithm}
\usepackage{algpseudocode}

\usepackage{colortbl}
\definecolor{lightgray}{rgb}{0.9,0.9,0.9}
\definecolor{lightred}{rgb}{1,0.8,0.8}
\definecolor{lightgreen}{rgb}{0.6,1,0.6}
\definecolor{lightyellow}{rgb}{1,1,0.5}
\definecolor{lightgrey}{rgb}{0.8,0.8,0.8}

\allowdisplaybreaks[1]

\usepackage{babel}
\providecommand{\conjecturename}{Conjecture}
\providecommand{\definitionname}{Definition}
\providecommand{\theoremname}{Theorem}

\makeatother

\usepackage{babel}
\providecommand{\conjecturename}{Conjecture}
\providecommand{\corollaryname}{Corollary}
\providecommand{\definitionname}{Definition}
\providecommand{\propositionname}{Proposition}
\providecommand{\theoremname}{Theorem}

\begin{document}
\title{Asymptotically Scale-invariant Multi-resolution Quantization}
\author{Cheuk Ting Li\thanks{This paper is the extended version of a paper
submitted to the IEEE International Symposium on Information Theory
2020.}\\
 Department of Information Engineering\\
 The Chinese University of Hong Kong\\
 Email: ctli@ie.cuhk.edu.hk}
\maketitle
\begin{abstract}
A multi-resolution quantizer is a sequence of quantizers where the
output of a coarser quantizer can be deduced from the output of a
finer quantizer. In this paper, we propose an asymptotically scale-invariant
multi-resolution quantizer, which performs uniformly across any choice
of average quantization step, when the length of the range of input
numbers is large. Scale invariance is especially useful in worst case
or adversarial settings, ensuring that the performance of the quantizer
would not be affected greatly by small changes of storage or error
requirements. We also show that the proposed quantizer achieves a
tradeoff between rate and error that is arbitrarily close to the optimum.
\end{abstract}

\section{Introduction}

A multi-resolution quantizer is a sequence of quantizers, where the
output of a coarser quantizer can be deduced from the output of a
finer quantizer (without knowledge of the original data). It has been
studied, for example, by Koshelev \cite{koshelev1980hierarchical},
Equitz and Cover \cite{equitz1991successive}, Rimoldi \cite{rimoldi1994successive},
Brunk and Farvardin \cite{brunk1996fixedrate}, Jafarkhani, Brunk,
and Farvardin \cite{brunk1997entropy}, Effros \cite{effros1998practical},
Wu and Dumitrescu \cite{wu2002multires,dumitrescu2004algorithms}
and Effros and Dugatkin \cite{effros2004multires}. There are two
main uses of multi-resolution quantizers: to allow a coarser quantization
to be obtained from a finer quantization by discarding some information,
and to allow a finer quantization to be obtained from a coarser quantization
by adding some additional information from the encoder (i.e., successive
refinement). We first focus on the first usage.

Consider the setting where a piece of data is relayed across a sequence
of nodes, where each communication link has a different capacity.
Each node only has information about the capacity of its incoming
and outgoing link, and therefore can only compress the incoming data
according to the capacity of the outgoing link and send it to the
next node, if the outgoing link has smaller capacity than the incoming
link (otherwise the node can relay the incoming data exactly). If
the data is a number $x\in[0,1)$, a simple scheme, which we call
the simple uniform quantizer, is that node $i$ would apply the uniform
quantization $y_{i+1}:=(\lfloor k_{i}y_{i}\rfloor+1/2)/k_{i}$ to
its incoming data $y_{i}$ and send it to the node $i+1$, where $k_{i}$
is the number of values that can be sent through the link between
node $i$ and $i+1$. This scheme is undesirable since if $k_{1}=4$,
$k_{2}=3$, and $x=y_{1}=2/7$, then $y_{2}=3/8$, $y_{3}=1/2$, giving
an absolute error $|y_{3}-x|=3/14$ that is larger than as if $k_{1}=k_{2}=3$
(i.e., the data is only compressed once according to the worse link),
giving an absolute error $5/42$.

To mitigate this problem, we can use a multi-resolution quantizer,
where the output of a coarser quantizer can be obtained from the output
of a finer quantizer, and thus the final output of the relay would
convey the same information as if the input is only compressed once
according to the link with the lowest capacity (by simply quantizing
the final output again according to the link with the lowest capacity).
One simple scheme, which we call the binary multi-resolution quantizer
(BMRQ), would be to quantize only using step sizes that are powers
of $2$, i.e., $y_{i+1}:=2^{-\lfloor\log_{2}k_{i}\rfloor}(\lfloor2^{\lfloor\log_{2}k_{i}\rfloor}y_{i}\rfloor+1/2)$.
For the aforementioned example $k_{1}=4$, $k_{2}=3$, $x=2/7$, we
have $y_{2}=3/8$, $y_{3}=1/4$, giving an absolute error $|y_{3}-x|=1/28$.

Nevertheless, the BMRQ does not perform well in worst case or adversarial
settings. Consider the setting where an adversary can modify $k_{i}$
to increase the quantization error. If $k_{1}=32$ (the quantization
step is $1/32$), then the adversary can reduce $k_{1}$ by $1$ to
$31$, increasing the quantization step two-fold to $1/16$ (and hence
the average absolute error is also increased two-fold). The BMRQ performs
well only when the $k_{i}$'s are powers of $2$. The adversary can
modify $k_{i}$ slightly off powers of $2$ to cause a significant
degradation of the quantized data. We call this problem scale dependence,
meaning that the multi-resolution quantizer does not perform uniformly
well for all choices of quantization step.

\begin{figure}
\begin{centering}
\includegraphics[scale=0.435]{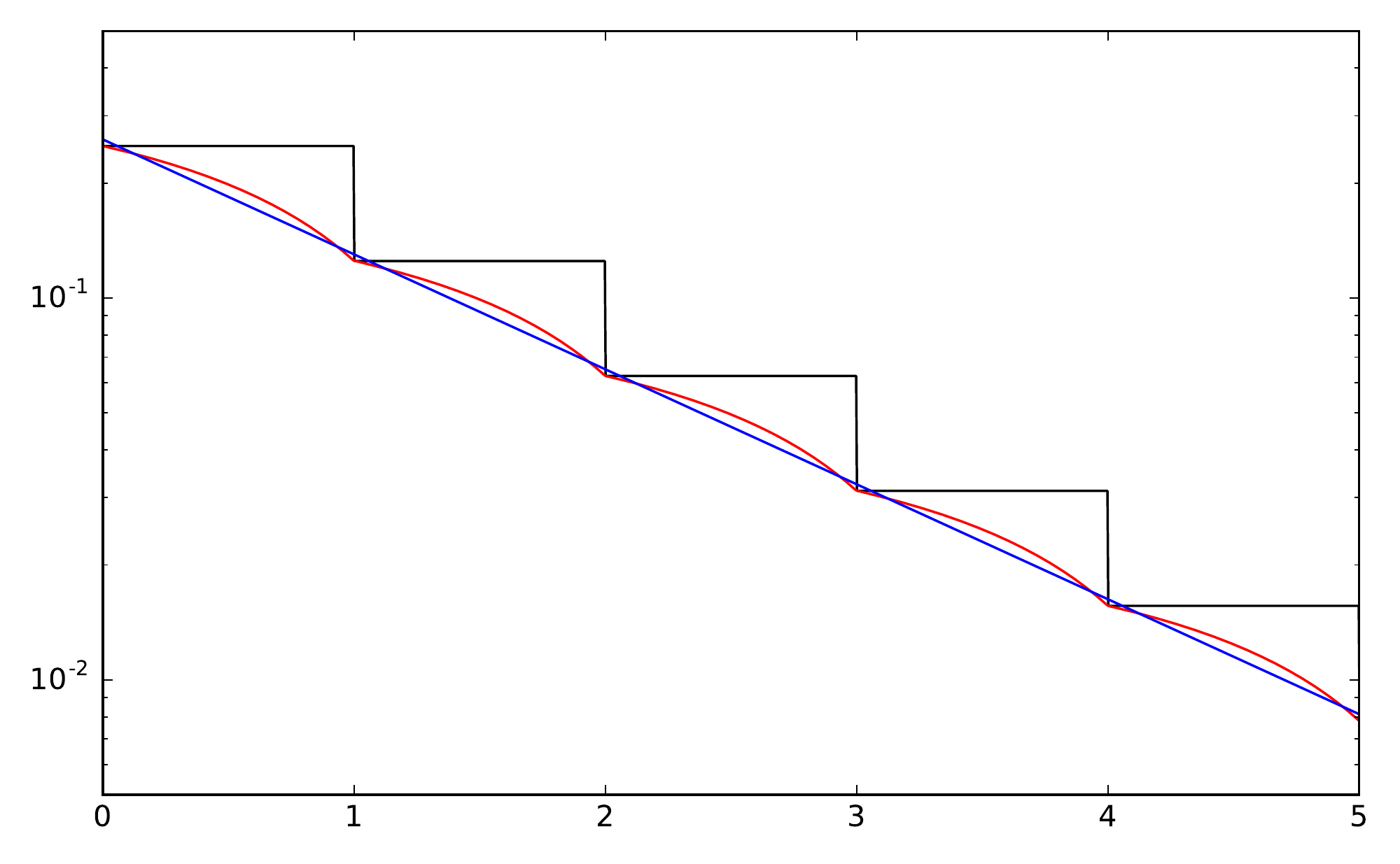} 
\par\end{centering}
\caption{\label{fig:bbvsb}Plot of the asymptotic $L^{1}$ error against the
log-rate (more precisely, plot of $\inf\{2^{-R_{2}(Q_{s})-2}:s>0,\,R_{0}(Q_{s})\le x\}$
against $x$) for the BMRQ $\{Q_{s}^{\mathrm{bin}}\}$ (black), the
DBMRQ $\{Q_{s}^{\mathrm{di}}\}$ (red), and the BBMRQ $\{Q_{s}^{\mathrm{bias},\alpha}\}$
for $\alpha$ close to $1/2$ (blue). We can see that the BBMRQ provides
a smoother trade-off between the error and the rate (the log plot
is a straight line due to scale invariance). Also, the BBMRQ outperforms
the BMRQ and DBMRQ except when the log rate is close to an integer.}
\end{figure}

In this paper, we introduce an asymptotically scale-invariant multi-resolution
quantizer, called the biased binary multi-resolution quantizer (BBMRQ),
that performs uniformly across any choice of average quantization
step (BBMRQ is a non-uniform quantizer), when the length of the range
of input numbers tends to infinity. Therefore its performance degrades
gracefully when the adversary modifies the communication constraints.
We show that the BBMRQ outperforms the BMRQ except when the average
quantization step is close to a power of $2$ (see Figure \ref{fig:bbvsb}).

Asymptotically scale-invariant multi-resolution quantizers are also
useful in successive refinement settings. Consider the scenario where
an encoder observes a number $x$ and produces a sequence of bits.
Due to storage or communication constraint, we only keep the first
$n$ bits and discard the rest, where $n$ is chosen according to
the storage constraint and the bit sequence. The BMRQ corresponds
to the scheme where the bit sequence is the binary representation
of $x$, and $n$ is chosen only according to the storage constraint.
For the BBMRQ, $n$ also depends on the bit sequence (i.e., it is
a variable length code), which allows the performance of the quantizer
to vary smoothly when the storage constraint changes. The bit sequence
can be produced according to the quantization tree (see Figure \ref{fig:bbtree}).

This paper is organized as follows. In Section \ref{sec:mrq}, we
give the criteria of multi-resolution quantizers and define the cell
size cdf for measuring the performance of a quantizer. In Section
\ref{sec:bbmrq}, we define the BBMRQ and present the main result
regarding the performance of BBMRQ. In Section \ref{sec:converse},
we show that the BBMRQ achieves a tradeoff between rate and error
that is arbitrarily close to the optimum.

\medskip{}

\subsection{Previous Work}

The seminal work by Equitz and Cover \cite{equitz1991successive}
concerns the problem of successive refinement of information, where
the lossy reconstruction is iteratively refined by supplying more
information. It is also studied by Rimoldi \cite{rimoldi1994successive}.
Also see \cite{koshelev1980hierarchical} for a related setting. Another
line of research is multiple description coding \cite{wolf1980source,ozarow1980source,gamal1982achievable},
where several descriptions are produced from the same source, and
the distortion of the reconstruction depends on which subset of descriptions
is available to the decoder. Also see \cite{yeung1995multilevel}
for a related setting. Note that the aforementioned papers concern
the asymptotic rate-distortion problem, whereas this paper focuses
on the one-shot scalar quantization setting.

Vaishampayan \cite{vaishampayan1993design} studied multiple description
scalar quantizers. Brunk and Farvardin \cite{brunk1996fixedrate}
and Jafarkhani, Brunk, and Farvardin \cite{brunk1997entropy} studied
multi-resolution scalar quantizers (or successively refinable quantizers),
and provide algorithms for designing quantizers with small error given
the distribution of the input. Effros \cite{effros1998practical}
and Effros and Dugatkin \cite{effros2004multires} studied multi-resolution
vector quantizers. Algorithms for multi-resolution quantization were
studied by Wu and Dumitrescu \cite{wu2002multires,dumitrescu2004algorithms}.
Note that the aforementioned papers concern the setting where the
distribution of the input is known, and the quantizers are designed
accordingly (as in the classical Lloyd-Max algorithm \cite{lloyd1982least,max1960quantizing}).
In this paper, we do not design the quantizer according to the input
distribution, but rather assume the input is (loosely speaking) uniform
over a long interval.

\medskip{}

\section{Multi-resolution Quantizer\label{sec:mrq}}

In this paper, quantizer can refer to any measurable function $Q:\mathbb{R}\to\mathbb{R}$,
where the range $Q(\mathbb{R})$ is a finite or countable set. We
call $Q$ a centered quantizer if each reconstruction level is the
center of its corresponding quantization cell, which is formally defined
below. 
\begin{defn}
We call a function $Q:\mathbb{R}\to\mathbb{R}$ a \emph{centered quantizer}
if it is non-decreasing, the range $Q(\mathbb{R})$ is a finite or
countable, and 
\begin{align*}
Q(x) & =\frac{1}{2}\Big(\inf\{y:Q(y)\!=\!Q(x)\}+\sup\{y:Q(y)\!=\!Q(x)\}\Big)
\end{align*}
for all $x\in\mathbb{R}$. 
\end{defn}
We give the criteria of multi-resolution quantizers below. 
\begin{defn}
We call $\{Q_{s}\}_{s>0}$ a \emph{multi-resolution quantizer (MRQ)
}if the functions $Q_{s}:\mathbb{R}\to\mathbb{R}$ are measurable
and satisfy 
\[
Q_{s_{2}}(Q_{s_{1}}(x))=Q_{s_{2}}(x)
\]
for any $s_{2}\ge s_{1}>0$ and $x\in\mathbb{R}$. The parameter $s$
(that can be any positive real number) usually corresponds to the
(maximum or average) quantization step. We call $\{Q_{s}\}_{s>0}$
a \emph{centered multi-resolution quantizer} if the functions $Q_{s}$
are centered quantizers.
\end{defn}
This definition ensures that $Q_{s_{2}}(x)$ can be computed using
$Q_{s_{1}}(x)$ for $s_{2}\ge s_{1}>0$, simply by quantizing $Q_{s_{1}}(x)$
using $Q_{s_{2}}$. As a result, if $s_{n}\ge s_{i}$ for all $i=1,\ldots,n$,
then $Q_{s_{n}}(Q_{s_{n-1}}(\cdots Q_{s_{1}}(x)\cdots))=Q_{s_{n}}(x)$,
i.e., we can quantize the final output of the relay again by the coarsest
quantizer to obtain a result the same as if the input is only compressed
once by the coarsest quantizer.

We remark that this definition is different from the previous definitions
(e.g. \cite{brunk1996fixedrate,brunk1997entropy,wu2002multires}),
which also concern how the quantized number is represented (e.g. by
a bit sequence). Here we only concern the mapping from the input number
to its reconstruction level, and assume that a suitable compression
algorithm is applied to the reconstruction levels if the multi-resolution
quantizer is to be used in practice.

Note that the simple uniform quantizer $Q_{s}(x):=s(\lfloor x/s\rfloor+1/2)$
is not a multi-resolution quantizer, since $Q_{3}(x)$ cannot be deduced
from $Q_{2}(x)$. Nevertheless, if we restrict the step size to powers
of $2$, i.e., $Q_{s}^{\mathrm{bin}}(x):=2^{\lfloor\log_{2}s\rfloor}(\lfloor2^{-\lfloor\log_{2}s\rfloor}x\rfloor+1/2)$,
then this is a MRQ, which we call the \emph{binary multi-resolution
quantizer} (BMRQ).

A downside of the BMRQ is that it is scale-dependent. The average
absolute error of the binary multi-resolution quantizer is $2^{\lfloor\log_{2}s\rfloor-2}$
(when the input is uniformly distributed over a long interval), which
must be a power of $2$. Therefore, the quantizer is suitable if the
maximum allowed average absolute error is a power of $2$, but not
suitable if the maximum allowed average absolute error is slightly
smaller than a power of $2$. Scale-dependence is particularly undesirable
in worst case or adversarial settings, where the quantizer must work
well for any maximum allowed average absolute error (or other error
metrics).

Loosely speaking, the BMRQ is the optimal uniform multi-resolution
quantizer (where for each $s$, $Q_{s}$ divides the real line into
intervals of the same length) \footnote{We can, for example, use step sizes that are powers of $3$, i.e.,
$Q_{s}(x):=3^{\lfloor\log_{3}s\rfloor}(\lfloor3^{-\lfloor\log_{3}s\rfloor}x\rfloor+1/2)$,
though it provides less control over the step size, since $\{3^{\lfloor\log_{3}s\rfloor}\}_{s}$
are spaced farther apart than $\{2^{\lfloor\log_{2}s\rfloor}\}_{s}$.}. Nevertheless, scale-dependence is an inherent disadvantage of uniform
multi-resolution quantizers. In order to overcome this disadvantage,
we consider non-uniform quantizers, where each quantization cell can
have different size. The distribution of cell sizes is captured by
the following definition. 
\begin{defn}
For a (not necessarily centered) quantizer $Q:\mathbb{R}\to\mathbb{R}$,
define its \emph{cell size cumulative distribution function (cell
size cdf) }on the measurable set $S\subseteq\mathbb{R}$ with positive
measure as 
\begin{align*}
F_{Q,S}(z) & :=\frac{\lambda\left(\left\{ x\in S:\,\lambda\left(\left\{ y\in S:\,Q(y)=Q(x)\right\} \right)\le z\right\} \right)}{\lambda(S)},
\end{align*}
where $\lambda$ denotes the Lebesgue measure. Define its \emph{asymptotic
cell size cdf }$F_{Q}$ to be the cdf that is the limit (with respect
to the Lévy metric) of $F_{Q,[x_{0},x_{1}]}$ as $x_{1}-x_{0}\to\infty$,
i.e., $F_{Q}$ is a cdf and 
\begin{equation}
\lim_{l\to\infty}\sup_{(x_{0},x_{1}):x_{1}-x_{0}\ge l}d_{\mathrm{L}}(F_{Q,[x_{0},x_{1}]},F_{Q})=0,\label{eq:levy_lim}
\end{equation}
where $d_{\mathrm{L}}(F,G):=\inf\{\epsilon>0:F(x-\epsilon)-\epsilon\le G(x)\le F(x+\epsilon)+\epsilon\,\forall\,x\in\mathbb{R}\}$
is the Lévy metric. Note that $F_{Q}$ may not exist for some $Q$.

\smallskip{}
 
\end{defn}
We then give the criteria for asymptotic scale invariance. 
\begin{defn}
For a multi-resolution quantizer $\{Q_{s}\}_{s>0}$, if $F_{Q_{s}}$
exists for any $s>0$, define 
\begin{align*}
\overline{F}_{\{Q_{s}\}_{s}}(x) & :=\sup_{s>0}\{F_{Q_{s}}(xs)\},\\
\text{\ensuremath{\underline{F}}}_{\{Q_{s}\}_{s}}(x) & :=\inf_{s>0}\{F_{Q_{s}}(xs)\}.
\end{align*}
We call $\{Q_{s}\}_{s>0}$ \emph{asymptotically scale-invariant} if
$\overline{F}_{\{Q_{s}\}_{s}}=\text{\ensuremath{\underline{F}}}_{\{Q_{s}\}_{s}}$,
i.e., the functions $x\mapsto F_{Q_{s}}(xs)$ are the same for all
$s>0$.

\smallskip{}
 
\end{defn}
One way to improve the BMRQ is to add more intermediate steps between
$Q_{2^{n}}^{\mathrm{bin}}$ and $Q_{2^{n+1}}^{\mathrm{bin}}$, where
only some of the adjacent pairs of quantization cells of $Q_{2^{n}}^{\mathrm{bin}}$
are merged. 
\begin{defn}
Define the \emph{dithered binary multi-resolution quantizer (DBMRQ)}
as 
\begin{align*}
Q_{s}^{\mathrm{di}}(x) & :=\!\!\begin{cases}
2^{\lfloor\log_{2}s\rfloor+1}(\lfloor2^{-\lfloor\log_{2}s\rfloor-1}x\rfloor\!+\!1/2)\\
\;\;\;\;\;\lefteqn{\mathrm{if}\,\mathrm{frac}(\phi\lfloor2^{-\lfloor\log_{2}s\rfloor-1}x\rfloor)<2\!-\!2^{\lfloor\log_{2}s\rfloor+1}/s}\\
2^{\lfloor\log_{2}s\rfloor}(\lfloor2^{-\lfloor\log_{2}s\rfloor}x\rfloor+1/2) & \mathrm{otherwise},
\end{cases}
\end{align*}
where $\mathrm{frac}(\gamma):=\gamma-\lfloor\gamma\rfloor$, and $\phi:=(1+\sqrt{5})/2$
is the golden ratio (or any irrational number works).

\smallskip{}
 
\end{defn}
See Figure \ref{fig:bbtree} for an illustration of DBMRQ. The quantizer
$Q_{s}^{\mathrm{di}}$ has two cell sizes: $2^{\lfloor\log_{2}s\rfloor}$
and $2^{\lfloor\log_{2}s\rfloor+1}$. The choice of which cell size
to use is determined by the $\mathrm{frac}$ function in the definition.
It can be checked that the cell size cdf of $Q_{s}^{\mathrm{di}}$
is 
\begin{align*}
F_{Q_{s}^{\mathrm{di}}}(x) & =(2-2^{\lfloor\log_{2}s\rfloor+1}/s)\mathbf{1}\{2^{(\lfloor\log_{2}s\rfloor+1)}\le x\}\\
 & \;\;\;+(2^{\lfloor\log_{2}s\rfloor+1}/s-1)\mathbf{1}\{2^{\lfloor\log_{2}s\rfloor}\le x\}.
\end{align*}
Note that the BMRQ and the DBMRQ are not asymptotically scale-invariant.

\section{Quantities of Interest}

The cell size cdf provides some information about the quantizer $Q$.
We define the following useful quantity.
\begin{defn}
For a quantizer $Q$ where $F_{Q}$ exists, define its \emph{Rényi
entropy rate} as
\[
R_{\eta}(Q):=\frac{1}{1-\eta}\log_{2}\int_{0}^{\infty}\gamma^{\eta-1}\mathrm{d}F_{Q}(\gamma)
\]
for $\eta\in\mathbb{R}_{>0}\backslash\{1\}$, and
\[
R_{1}(Q):=\int_{0}^{\infty}\log_{2}(\gamma^{-1})\mathrm{d}F_{Q}(\gamma).
\]
We call $R_{0}(Q)$ the \emph{log-rate} of $Q$.\smallskip{}
\end{defn}
Several quanities of interest can be obtained from the Rényi entropy
rate. If $X\sim\mathrm{Unif}[x_{0},x_{1}]$, then:
\begin{itemize}
\item The number of reconstruction levels (possible values of $Q(X)$ with
positive probability) is 
\[
|\{x:\,\mathbf{P}(Q(X)=x)>0\}|=(x_{1}-x_{0})\int_{0}^{\infty}\gamma^{-1}\mathrm{d}F_{Q,[x_{0},x_{1}]}(\gamma).
\]
The reason is that for a quantization cell of size $\gamma$, the
probability that $X$ is in that cell is $\gamma/(x_{1}-x_{0})$,
and hence its contribution to $(x_{1}-x_{0})\int_{0}^{\infty}\gamma^{-1}\mathrm{d}F_{Q,[x_{0},x_{1}]}(\gamma)$
is $(x_{1}-x_{0})\gamma^{-1}(\gamma/(x_{1}-x_{0}))=1$. Hence, the
number of bits needed to encode the levels (using fixed-length code)
is $\lceil\log_{2}((x_{1}-x_{0})\int_{0}^{\infty}\gamma^{-1}\mathrm{d}F_{Q,[x_{0},x_{1}]}(\gamma))\rceil$.
Therefore, the log-rate
\[
R_{0}(Q)=\log_{2}\int_{0}^{\infty}\gamma^{-1}\mathrm{d}F_{Q}(\gamma)
\]
is the logarithm of the rate of increase of the number of reconstruction
levels as the interval $[x_{0},x_{1}]$ becomes longer.\smallskip{}
\item The entropy of the output is 
\[
H(Q(X))=\int_{0}^{\infty}\log_{2}(\gamma^{-1}(x_{1}-x_{0}))\mathrm{d}F_{Q,[x_{0},x_{1}]}(\gamma).
\]
Therefore,
\[
R_{1}(Q)=\int_{0}^{\infty}\log_{2}(\gamma^{-1})\mathrm{d}F_{Q}(\gamma)
\]
describes how $H(Q(X))$ increases as the interval $[x_{0},x_{1}]$
becomes longer.\smallskip{}
\item The average $L^{p}$ error is lower-bounded by 
\begin{equation}
\mathbf{E}[|X\!-\!Q(X)|^{p}]\ge\int_{0}^{\infty}\frac{(\gamma/2)^{p}}{p+1}\mathrm{d}F_{Q,[x_{0},x_{1}]}(\gamma).\label{eq:lp_finite}
\end{equation}
The reason is that for a quantization cell of size $\gamma$, the
expected $L^{p}$ error conditioned on that $X$ is in that cell is
at least $(\gamma/2)^{p}/(p+1)$ (equality holds if the quantization
cell is an interval, and the reconstruction level is its midpoint).
The following proposition shows the relation between the asymptotic
$L^{p}$ error and $R_{p+1}(Q)$.
\end{itemize}
\begin{prop}
\label{prop:lp_asymp}Fix $p>0$ and a quantizer $Q$ where $F_{Q}$
exists. Let $X\sim\mathrm{Unif}[x_{0},x_{1}]$. We have
\[
\underset{x_{1}-x_{0}\to\infty}{\lim\inf}\mathbf{E}[|X-Q(X)|^{p}]\ge\frac{1}{p+1}2^{-p(R_{p+1}(Q)+1)}.
\]
Moreover, if $Q$ is centered, then
\[
\lim_{x_{1}-x_{0}\to\infty}\mathbf{E}[|X-Q(X)|^{p}]=\frac{1}{p+1}2^{-p(R_{p+1}(Q)+1)}.
\]
\end{prop}
\begin{IEEEproof}
For the first part, if $x_{1}-x_{0}\to\infty$, then $F_{Q,[x_{0},x_{1}]}\to F_{Q}$
(in Lévy metric), and hence by \eqref{eq:lp_finite},
\begin{align*}
 & \underset{x_{1}-x_{0}\to\infty}{\lim\inf}\mathbf{E}[|X\!-\!Q(X)|^{p}]\\
 & \ge\underset{x_{1}-x_{0}\to\infty}{\lim\inf}\int_{0}^{\infty}\frac{(\gamma/2)^{p}}{p+1}\mathrm{d}F_{Q,[x_{0},x_{1}]}(\gamma)\\
 & \ge\int_{0}^{\infty}\frac{(\gamma/2)^{p}}{p+1}\mathrm{d}F_{Q}(\gamma)\\
 & =\frac{1}{p+1}2^{-p(R_{p+1}(Q)+1)}.
\end{align*}
For the second part, assume that $Q$ is centered. We first show that
there exists $b>0$ such that each quantization cell has size upper-bounded
by $b$. Assume the contrary that the quantization cells can be arbitrarily
large. Then there exists a sequence $\{(x_{0,i},x_{1,i})\}_{i}$ such
that $x_{1,i}-x_{0,i}\to\infty$ and $F_{Q,[x_{0,i},x_{1,i}]}(\gamma)=\mathbf{1}\{\gamma\ge x_{1,i}-x_{0,i}\}$
(take $x_{0,i},x_{1,i}$ to be the end points of cells in a sequence
of cells with sizes tend to $\infty$). This contradicts \eqref{eq:levy_lim}
since $F_{Q,[x_{0,i},x_{1,i}]}$ does not have a limit. Hence such
$b>0$ exists. For $X\sim\mathrm{Unif}[x_{0},x_{1}]$, we have
\begin{align*}
 & \mathbf{E}[|X-Q(X)|^{p}]\\
 & \le\mathbf{E}\left[\mathbf{1}\{X\in[x_{0}+b,\,x_{1}-b]\}|X-Q(X)|^{p}\right]+\mathbf{P}\left(X\notin[x_{0}+b,\,x_{1}-b]\right)\frac{(b/2)^{p}}{p+1}\\
 & \le\int_{0}^{\infty}\frac{(\gamma/2)^{p}}{p+1}\mathrm{d}F_{Q,[x_{0},x_{1}]}(\gamma)+\frac{2b}{x_{1}-x_{0}}\cdot\frac{(b/2)^{p}}{p+1}\\
 & \to\int_{0}^{\infty}\frac{(\gamma/2)^{p}}{p+1}\mathrm{d}F_{Q}(\gamma)\,=\,\frac{1}{p+1}2^{-p(R_{p+1}(Q)+1)}
\end{align*}
as $x_{1}-x_{0}\to\infty$.
\end{IEEEproof}
\[
\]

\section{Biased Binary Multi-resolution Quantizer\label{sec:bbmrq}}

We now state the main result in this paper, which is proved later
in this section. 
\begin{thm}
\label{thm:opt_approx}For any $\epsilon>0$, there exists an asymptotically
scale-invariant centered MRQ $\{Q_{s}\}_{s>0}$ with $\overline{F}_{\{Q_{s}\}_{s}}(1/2-\epsilon)=0$,
$\overline{F}_{\{Q_{s}\}_{s}}(1)=1$, and $d_{\mathrm{L}}(\overline{F}_{\{Q_{s}\}_{s}},F_{2^{\mathrm{Unif}[-1,0]}})<\epsilon$,
where 
\[
F_{2^{\mathrm{Unif}[-1,0]}}(x):=\min\left\{ \max\{\log_{2}(x)+1,\,0\},\,1\right\} ,
\]
i.e., $F_{2^{\mathrm{Unif}[-1,0]}}$ is the cdf of $2^{Z}$ where
$Z\sim\mathrm{Unif}[-1,0]$.

\medskip{}
\end{thm}
As a result, the Rényi entropy rate $R_{\eta}(Q_{s})$ can be arbitrarily
close to
\begin{equation}
\frac{1}{1-\eta}\log_{2}\left(\frac{1-2^{1-\eta}}{\eta-1}\log_{2}e\right)-\log_{2}s\label{eq:opt_approx_renyi}
\end{equation}
for $\eta\neq1$, and $R_{1}(Q_{s})$ can be arbitrarily close to
$1/2-\log_{2}s$. We will show in Corollary \ref{cor:rate_error}
that this tradeoff between log-rate and $L^{p}$ error can be arbitrarily
close to optimal.

\begin{figure*}
\begin{centering}
\includegraphics[scale=0.44]{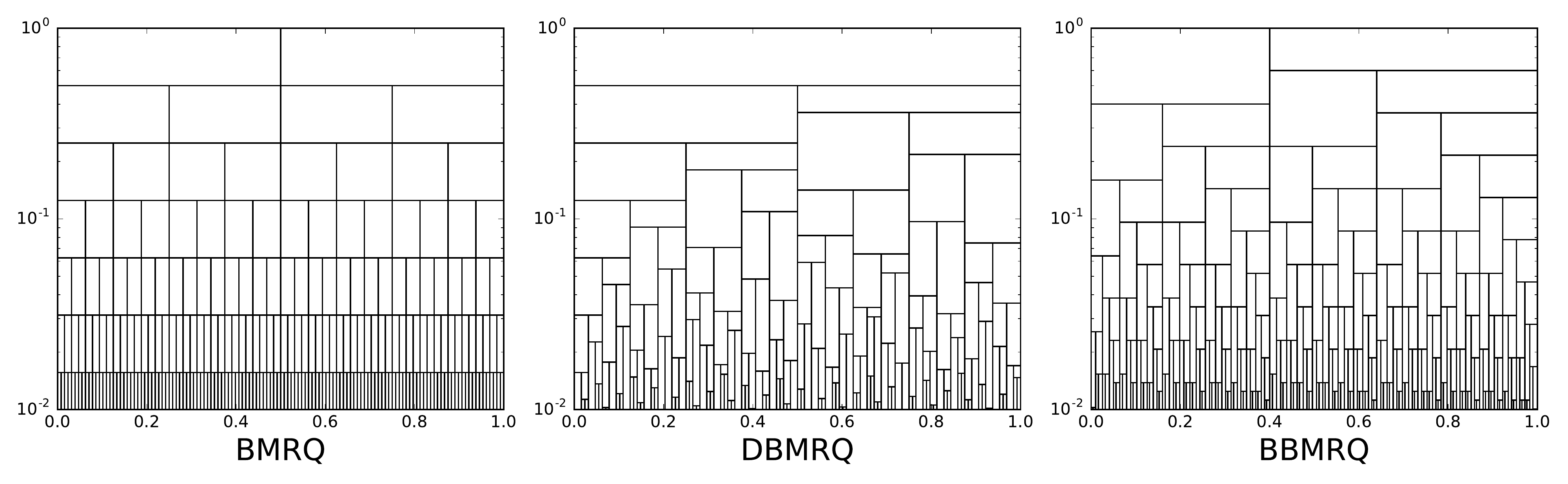} 
\par\end{centering}
\caption{\label{fig:bbtree}The quantization tree of BMRQ, DBMRQ and BBMRQ
(from left to right). The x-axis is the input $x$, and the y-axis
is the parameter $s$. A rectangle in the figure means that each point
$(x,s)$ in the rectangle has the same quantized value $Q_{s}(x)$.}
\end{figure*}

To prove Theorem \ref{thm:opt_approx}, we introduce the following
construction. 
\begin{defn}
We define the \emph{biased quantization tree }with parameter $0<\alpha<1$
recursively as follows. For any $n\in\mathbb{Z}$ and sequence $\{z_{i}\}_{i\le n}$
($z_{i}\in\{0,1\}$, the index $i$ is over $\mathbb{Z}\cap(-\infty,n]$)
with finitely many $1$'s, define 
\begin{align*}
 & (\text{\ensuremath{\underline{T}}}_{n}^{\mathrm{bias},\alpha}(\{z_{i}\}_{i\le n}),\overline{T}_{n}^{\mathrm{bias},\alpha}(\{z_{i}\}_{i\le n}))\\
 & \!:=\!\begin{cases}
\left(0,\,\alpha^{n}\right) & \!\!\!\!\!\!\!\!\!\!\!\!\!\!\!\!\!\!\mathrm{if}\,z_{n}\!=\!z_{n-1}\!=\!\cdots\!=\!0\\
\left(\text{\ensuremath{\underline{T}}}_{n-1},\alpha\overline{T}_{n-1}\!+\!(1\!-\!\alpha)\text{\ensuremath{\underline{T}}}_{n-1}\right) & \mathrm{else}\,\mathrm{if}\;z_{n}=0\\
\left(\alpha\overline{T}_{n-1}\!+\!(1\!-\!\alpha)\text{\ensuremath{\underline{T}}}_{n-1},\overline{T}_{n-1}\right) & \mathrm{otherwise}.
\end{cases}
\end{align*}
where we write $\overline{T}_{n-1}=\overline{T}_{n-1}^{\mathrm{bias},\alpha}(\{z_{i}\}_{i\le n-1})$
(likewise for $\text{\ensuremath{\underline{T}}}_{n-1}$) for brevity.
Note that the first case above serves as the base case of the recursive
definition.

Define the \emph{biased binary multi-resolution quantizer (BBMRQ)
}with parameter $0<\alpha<1$ as follows. For $x\ge0$, define 
\begin{align*}
Q_{s}^{\mathrm{bias},\alpha}(x) & :=\frac{1}{2}\left(\text{\ensuremath{\underline{T}}}_{n}^{\mathrm{bias},\alpha}(\{z_{i}\}_{i\le n})+\overline{T}_{n}^{\mathrm{bias},\alpha}(\{z_{i}\}_{i\le n})\right),
\end{align*}
where $\{z_{i}\}_{i\le n}$ satisfies $x\in[\text{\ensuremath{\underline{T}}}_{n}^{\mathrm{bias},\alpha}(\{z_{i}\}_{i\le n}),\overline{T}_{n}^{\mathrm{bias},\alpha}(\{z_{i}\}_{i\le n}))$
and $\overline{T}_{n}^{\mathrm{bias},\alpha}(\{z_{i}\}_{i\le n})-\text{\ensuremath{\underline{T}}}_{n}^{\mathrm{bias},\alpha}(\{z_{i}\}_{i\le n})\le s$,
and we select the $\{z_{i}\}_{i\le n}$ with the smallest $n$ satisfying
these two constraints. For $x<0$, define $Q_{s}^{\mathrm{bias},\alpha}(x):=-Q_{s}^{\mathrm{bias},\alpha}(-x)$.

\medskip{}
 
\end{defn}
Intuitively, the BBMRQ repeatedly divides an interval into two subintervals
of proportion $\alpha$ and $1-\alpha$, until the lengths of the
intervals fall below $s$ (see Figure \ref{fig:bbtree}). It is clear
that the BBMRQ is a centered MRQ.

We then use the BBMRQ to prove Theorem \ref{thm:opt_approx}. 
\begin{IEEEproof}
Fix any $1/2<\alpha<3/4$ such that $(\ln\alpha)/(\ln(1-\alpha))$
is irrational. Write $Q_{s}=Q_{s}^{\mathrm{bias},\alpha}$ for brevity.
Let $X\sim\mathrm{Unif}[0,1]$, and 
\[
W_{t}:=2^{t}\lambda\left(\left\{ y:\,Q_{2^{-t}}(y)=Q_{2^{-t}}(X)\right\} \right)
\]
for $t\ge0$, where $\lambda$ denotes the Lebesgue measure. Note
that $-\log_{2}W_{t}$ is the residual life of a renewal process with
interarrival time distribution $\alpha\delta_{-\log_{2}\alpha}+(1-\alpha)\delta_{-\log_{2}(1-\alpha)}$,
where $\delta_{w}$ is the degenerate distribution at $w$ (as $t$
increases, each time the quantization cell containing $X$ splits
into two, there is a probability $\alpha$ for $X$ to be in the cell
with proportion $\alpha$, and a probability $1-\alpha$ for $X$
to be in the cell with proportion $1-\alpha$). Since $(\ln\alpha)/(\ln(1-\alpha))$
is irrational, the interarrival times have a non-lattice distribution.
Denote the Markov kernel (conditional distribution of $W_{s+t}$ given
$W_{s}$) as $\kappa_{\alpha}^{t}$. By the key renewal theorem \cite{ross1995stochastic},
we have 
\[
d_{\mathrm{L}}(\kappa_{\alpha}^{t}\delta_{1},F_{\mathrm{bias},\alpha})\to0
\]
as $t\to\infty$, where $\kappa_{\alpha}^{t}\delta_{1}$ denotes the
distribution of $W_{t}$ conditioned on $W_{0}=1$, and $F_{\mathrm{bias},\alpha}$
is the cdf of the stationary distribution of the process $\{W_{t}\}$,
given by 
\begin{align}
F_{\mathrm{bias},\alpha}(\gamma) & :=\left(-\alpha\log_{2}\alpha-(1-\alpha)\log_{2}(1-\alpha)\right)^{-1}\nonumber \\
 & \;\;\;\cdot\Big(\alpha\max\left\{ \log_{2}(\min\{\gamma,1\}/\alpha),0\right\} \nonumber \\
 & \;\;\;\;\;+(1\!-\!\alpha)\max\left\{ \log_{2}(\min\{\gamma,1\}/(1\!-\!\alpha)),0\right\} \!\Big).\label{eq:fbias}
\end{align}
For $\tau>0$, let $\epsilon_{\tau}$ be such that $d_{\mathrm{L}}(\kappa_{\alpha}^{t}\delta_{1},F_{\mathrm{bias},\alpha})\le\epsilon_{\tau}$
for all $t\ge\tau$, and $\epsilon_{\tau}\to0$ as $\tau\to\infty$.

Fix any $s>0$, $\tau>1$ and $l>\tau2^{\tau+3}s$. Fix any $x_{0},x_{1}$
such that $x_{1}-x_{0}\ge l$. Let $X\sim\mathrm{Unif}[x_{0},x_{1}]$.
Let $[\text{\ensuremath{\underline{T}}}_{n^{(i)}}(z^{(i)}),\overline{T}_{n^{(i)}}(z^{(i)}))$
be the quantization cells of $Q_{2^{\tau+2}s}$ that are subsets of
$[x_{0},x_{1}]$, where $n^{(i)}\in\mathbb{Z}$ and $z^{(i)}\in\{0,1\}^{\mathbb{Z}\cap(-\infty,n^{(i)}]}$
for $i=1,2,\ldots,\tilde{N}$. Since $X$ is not in one of $[\text{\ensuremath{\underline{T}}}_{n^{(i)}}(z^{(i)}),\overline{T}_{n^{(i)}}(z^{(i)}))$
only when $X$ is in a quantization cell of $Q_{2^{\tau+2}s}$ that
includes $x_{0}$ or $x_{1}$, and each quantization cell of $Q_{2^{\tau+2}s}$
has length between $2^{\tau}s$ and $2^{\tau+2}s$ (since $1/2<\alpha<3/4$),
we have 
\begin{align*}
\mathbf{P}\bigg(\!X\notin\bigcup_{i=1}^{\tilde{N}}[\text{\ensuremath{\underline{T}}}_{n^{(i)}}(z^{(i)}),\overline{T}_{n^{(i)}}(z^{(i)}))\!\bigg) & \le\frac{2(2^{\tau+2}s)}{x_{1}-x_{0}}\\
 & \le1/\tau.
\end{align*}
Let $W_{t}:=2^{t}\lambda(\{y:\,Q_{2^{-t}}(y)=Q_{2^{-t}}(X)\})$. Conditioned
on $X\in[\text{\ensuremath{\underline{T}}}_{n^{(i)}}(z^{(i)}),\overline{T}_{n^{(i)}}(z^{(i)}))$
(let $L:=\overline{T}_{n^{(i)}}(z^{(i)})-\text{\ensuremath{\underline{T}}}_{n^{(i)}}(z^{(i)})$),
we have $W_{-\log_{2}L}=1$, and $\{W_{t}\}_{t\ge-\log_{2}L}$ is
a stochastic process with Markov kernel $\kappa_{\alpha}^{t}$, and
thus $W_{-\log_{2}s}\sim\kappa_{\alpha}^{\log_{2}L-\log_{2}s}\delta_{1}$,
and $d_{\mathrm{L}}(\kappa_{\alpha}^{\log_{2}L-\log_{2}s}\delta_{1},F_{\mathrm{bias},\alpha})\le\epsilon_{\tau}$
since $\log_{2}L-\log_{2}s\ge\log_{2}2^{\tau}s-\log_{2}s=\tau$. Let
$\bar{W}:=s^{-1}\lambda(\{y\in[x_{0},x_{1}]:\,Q_{s}(y)=Q_{s}(x)\})$.
Since $\bar{W}=W_{-\log_{2}s}$ if $X\in\bigcup_{i=1}^{\tilde{N}}[\text{\ensuremath{\underline{T}}}_{n^{(i)}}(z^{(i)}),\overline{T}_{n^{(i)}}(z^{(i)}))$,
we have $d_{\mathrm{L}}(F_{\bar{W}},F_{\mathrm{bias},\alpha})<\epsilon_{\tau}+1/\tau$.
Therefore $F_{\bar{W}}\to F_{\mathrm{bias},\alpha}$ as $\tau\to\infty$,
and thus $\{Q_{s}\}$ is asymptotically scale-invariant with cell
size cdf $F_{\mathrm{bias},\alpha}$. The result follows from the
fact that $d_{\mathrm{L}}(F_{\mathrm{bias},\alpha},F_{2^{\mathrm{Unif}[-1,0]}})\to0$
as $\alpha\to1/2$. 
\end{IEEEproof}
\medskip{}
\medskip{}

\section{Converse Results\label{sec:converse}}

In this section, we show a fundamental tradeoff between the log-rate
and the $L^{p}$ error of asymptotically scale-invariant multi-resolution
quantizers, and that BBMRQ can be arbitrarily close to optimal in
this regard. We first show an inequality that must be satisfied by
all asymptotically scale-invariant multi-resolution quantizers.
\begin{thm}
\label{thm:pos_neg}For any asymptotically scale-invariant multi-resolution
quantizer $\{Q_{s}\}_{s>0}$ with finite log-rate (i.e., $R_{0}(Q_{1})=\log_{2}\int_{0}^{\infty}\gamma^{-1}\mathrm{d}\overline{F}(\gamma)<\infty$,
where we write $\overline{F}=\overline{F}_{\{Q_{s}\}_{s}}$), the
distribution given by the cdf $\overline{F}$ is a continuous distribution,
and its pdf $\bar{f}$ satisfies
\begin{equation}
\int_{0}^{\infty}x^{-1}\left(1+\mathbf{1}\left\{ \overline{f}(x)\ge\zeta\overline{f}(x\zeta)\right\} \right)\left(\overline{f}(x)-\zeta\overline{f}(x\zeta)\right)\mathrm{d}x\le0\label{eq:thm_bd1}
\end{equation}
for any $\zeta>1$. As a result,
\begin{equation}
\bar{f}(y)\le\int_{0}^{\infty}x^{-1}\left(1+\mathbf{1}\{x>y\}\right)\bar{f}(x)\mathrm{d}x\label{eq:thm_bd2}
\end{equation}
for almost all $y>0$.
\end{thm}
\begin{IEEEproof}
Let $p_{s,\gamma}$ be the probability mass function corresponding
to the cdf $F_{Q_{s},[-\gamma,\gamma]}$ for $\gamma>0$ (note that
it is a discrete distribution so pmf exists). The number of quantization
cells of $Q_{s}$ in $[-\gamma,\gamma]$ of size $x$ is $2\gamma x^{-1}p_{s,\gamma}(x)$.
Let $n_{x}$ (resp. $\tilde{n}_{x}$) be the number of quantization
cells of $Q_{s}$ (resp. $Q_{s/\gamma}$) in $[-\gamma,\gamma]$ of
size $x$ that are not quantization cells of $Q_{s/\gamma}$ (resp.
$Q_{s}$). We have $n_{x}-\tilde{n}_{x}=2\gamma x^{-1}(p_{s,\gamma}(x)-p_{s/\zeta,\gamma}(x))$.
Since each cell of $Q_{s}$ is split into 2 or more cells in $Q_{s/\gamma}$,
we have $2\sum_{x}n_{x}\le\sum_{x}\tilde{n}_{x}$ (the summation is
over $x$ in the support of $p_{s,\gamma}$ or $p_{s/\zeta,\gamma}$),
and hence
\begin{align}
 & 2\sum_{x:\,p_{s,\gamma}(x)>p_{s/\zeta,\gamma}(x)}x^{-1}(p_{s,\gamma}(x)-p_{s/\zeta,\gamma}(x))\nonumber \\
 & =2\gamma^{-1}\sum_{x:\,n_{x}>\tilde{n}_{x}}(n_{x}-\tilde{n}_{x})\nonumber \\
 & =\gamma^{-1}\left(2\sum_{x}n_{x}-2\sum_{x}\min\{n_{x},\tilde{n}_{x}\}\right)\nonumber \\
 & \le\gamma^{-1}\left(\sum_{x}\tilde{n}_{x}-2\sum_{x}\min\{n_{x},\tilde{n}_{x}\}\right)\nonumber \\
 & \le\gamma^{-1}\sum_{x:\,n_{x}<\tilde{n}_{x}}(\tilde{n}_{x}-n_{x})\nonumber \\
 & =\sum_{x:\,p_{s,\gamma}(x)<p_{s/\zeta,\gamma}(x)}x^{-1}(p_{s/\zeta,\gamma}(x)-p_{s,\gamma}(x)).\label{eq:psgamma_bd}
\end{align}

Let $\nu$ be the signed measure induced by $\overline{F}(x)-\overline{F}(x\zeta)$
(i.e., $\nu(B)=\int_{B}\mathrm{d}(\overline{F}(x)-\overline{F}(x\zeta))$).
Fix any $\epsilon>0$. Let $A\subseteq\mathbb{R}_{>0}$ be a measurable
set. Since the measure $B\mapsto\int_{B}x^{-1}\mathrm{d}(\overline{F}(x)+\overline{F}(x\zeta))$
is a regular measure over $\mathbb{R}_{>0}$ \footnote{We can reparameterize the space by $y=\ln x$. Then $\int_{B}x^{-1}\mathrm{d}(\overline{F}(x)+\overline{F}(x\zeta))=\int_{\ln(B)}e^{-y}\mathrm{d}(\overline{F}(e^{y})+\overline{F}(e^{y}\zeta))$.
We have $\int_{B'}e^{-y}\mathrm{d}(\overline{F}(e^{y})+\overline{F}(e^{y}\zeta))\le2e^{-\inf B'}$,
and hence the measure $B'\mapsto\int_{B'}e^{-y}\mathrm{d}(\overline{F}(e^{y})+\overline{F}(e^{y}\zeta))$
is locally finite, and hence regular.}, there exists an open set $A'$ such that $A\subseteq A'$ and $\int_{A'\backslash A}x^{-1}\mathrm{d}(\overline{F}(x)+\overline{F}(x\zeta))<\epsilon$.
Since $A'$ is open, it can be expressed as a countable disjoint union
of open intervals $A'=\bigcup_{i}(a_{i},b_{i})$. Let $A'_{\delta}:=\bigcup_{i}(a_{i}+\delta,b_{i}-\delta)$
for $\delta>0$ (where $(a_{i}+\delta,b_{i}-\delta)=\emptyset$ if
$a_{i}+\delta\ge b_{i}-\delta$). Let $g_{\delta}(x):=\min\{(x-a_{i})\delta^{-1},\,(b_{i}-x)\delta^{-1},\,1\}$
if $x\in(a_{i},b_{i})$ ($g_{\delta}(x)=0$ if $x$ is not in any
of those invervals). Then $g_{\delta}$ is continuous and $\mathbf{1}\{x\in A'_{\delta}\}\le g_{\delta}(x)\le\mathbf{1}\{x\in A'\}$.
Hence,
\begin{align*}
 & \int_{A}x^{-1}\nu(\mathrm{d}x)\\
 & \le\int_{A'}x^{-1}\nu(\mathrm{d}x)+\int_{A'\backslash A}x^{-1}\mathrm{d}(\overline{F}(x)+\overline{F}(x\zeta))\\
 & \le\int x^{-1}g_{\delta}(x)\nu(\mathrm{d}x)+\int_{A'\backslash A'_{\delta}}x^{-1}\mathrm{d}(\overline{F}(x)+\overline{F}(x\zeta))+\epsilon\\
 & \to\int x^{-1}g_{\delta}(x)\nu(\mathrm{d}x)+\epsilon
\end{align*}
as $\delta\to0$ since $A'_{\delta}\nearrow A'$. Also, 
\begin{align*}
 & \int_{A}x^{-1}\nu(\mathrm{d}x)\\
 & \ge\int_{A'}x^{-1}\nu(\mathrm{d}x)\\
 & \ge\int x^{-1}g_{\delta}(x)\nu(\mathrm{d}x)-\int_{A'\backslash A'_{\delta}}x^{-1}\mathrm{d}(\overline{F}(x)+\overline{F}(x\zeta))\\
 & \to\int x^{-1}g_{\delta}(x)\nu(\mathrm{d}x)
\end{align*}
as $\delta\to0$. Therefore, we can let $g:=g_{\delta}$ for a $\delta>0$
small enough that 
\begin{equation}
\left|\int_{A}x^{-1}\nu(\mathrm{d}x)-\int x^{-1}g(x)\nu(\mathrm{d}x)\right|<2\epsilon.\label{eq:int_gdelta}
\end{equation}
Since $F_{Q_{s},[-\gamma,\gamma]}\to\overline{F}$ as $\gamma\to\infty$,
and $x\mapsto1+g(x)$ is bounded and continuous, we have $\int x^{-1}(1+g(x))\mathrm{d}F_{Q_{1},[-\gamma,\gamma]}(x)\to\int x^{-1}(1+g(x))\mathrm{d}\overline{F}(x)$.
Also, $\int x^{-1}(1+g(x))\mathrm{d}F_{Q_{1/\zeta},[-\gamma,\gamma]}(x)\to\int x^{-1}(1+g(x))\mathrm{d}\overline{F}(x\zeta)$.
As a result,
\begin{align}
 & \int x^{-1}(1+g(x))\mathrm{d}\big(F_{Q_{1},[-\gamma,\gamma]}(x)-F_{Q_{1/\zeta},[-\gamma,\gamma]}(x)\big)\nonumber \\
 & \to\int x^{-1}(1+g(x))\mathrm{d}\big(\overline{F}(x)-\overline{F}(x\zeta)\big)\nonumber \\
 & =\int x^{-1}(1+g(x))\nu(\mathrm{d}x).\label{eq:gp1_lim}
\end{align}
It can be deduced from \eqref{eq:psgamma_bd} that 
\[
\sum_{x}x^{-1}(1+\tilde{g}(x))(p_{s,\gamma}(x)-p_{s/\zeta,\gamma}(x))\le0
\]
for any $\tilde{g}:\mathbb{R}_{>0}\to[0,1]$, where the summation
is over $x$ in the support of $p_{s,\gamma}$ or $p_{s/\zeta,\gamma}$
(because to maximize the above expression, we should assign $\tilde{g}(x)=0$
to $x$'s where $p_{s/\zeta,\gamma}(x)-p_{s,\gamma}(x)>0$, and $\tilde{g}(x)=1$
otherwise). Substituting $\tilde{g}=g$,
\[
\int x^{-1}(1+g(x))\mathrm{d}\big(F_{Q_{1},[-\gamma,\gamma]}(x)-F_{Q_{1/\zeta},[-\gamma,\gamma]}(x)\big)\le0.
\]
By \eqref{eq:gp1_lim},
\[
\int x^{-1}(1+g(x))\nu(\mathrm{d}x)\le0.
\]
By \eqref{eq:int_gdelta},
\[
\int x^{-1}\nu(\mathrm{d}x)+\int_{A}x^{-1}\nu(\mathrm{d}x)\le2\epsilon.
\]
Taking $\epsilon\to0$ and rearranging the terms,
\begin{equation}
\int x^{-1}(1+\mathbf{1}\{x\in A\})\nu(\mathrm{d}x)\le0,\label{eq:bd_A_nu}
\end{equation}
\[
\int_{0}^{\infty}x^{-1}(1+\mathbf{1}\{x\in A\})\mathrm{d}\overline{F}(x)\le\zeta\int_{0}^{\infty}x^{-1}(1+\mathbf{1}\{x/\zeta\in A\})\mathrm{d}\overline{F}(x),
\]
\[
\int_{0}^{\infty}x^{-1}(1+\mathbf{1}\{x\in A\}-\zeta-\zeta\mathbf{1}\{x/\zeta\in A\})\mathrm{d}\overline{F}(x)\le0.
\]
Substituting $A=\{x:\,x>y\}$,
\begin{align*}
0 & \ge\int_{0}^{\infty}x^{-1}(1+\mathbf{1}\{x>y\}-\zeta-\zeta\mathbf{1}\{x>y\zeta\})\mathrm{d}\overline{F}(x)\\
 & =\int_{0}^{\infty}x^{-1}\left(\mathbf{1}\{y<x\le y\zeta\}-(\zeta-1)\left(1+\mathbf{1}\{x>y\zeta\}\right)\right)\mathrm{d}\overline{F}(x)\\
 & \ge\frac{\overline{F}(y\zeta)-\overline{F}(y)}{y\zeta}-(\zeta-1)\int_{0}^{\infty}x^{-1}\left(1+\mathbf{1}\{x>y\zeta\}\right)\mathrm{d}\overline{F}(x).
\end{align*}
Hence,
\[
\frac{\overline{F}(y\zeta)-\overline{F}(y)}{y\zeta-y}\le\zeta\int_{0}^{\infty}x^{-1}\left(1+\mathbf{1}\{x>y\zeta\}\right)\mathrm{d}\overline{F}(x).
\]
Since the right hand side is bounded by $\zeta2^{R_{0}+1}$, $\bar{f}$
exists and \eqref{eq:thm_bd2} holds for almost all $y>0$. Therefore,
\eqref{eq:bd_A_nu} becomes
\[
\int x^{-1}(1+\mathbf{1}\{x\in A\})\left(\overline{f}(x)-\zeta\overline{f}(x\zeta)\right)dx\le0.
\]
We can obtain \eqref{eq:thm_bd1} by substituting $A=\{x:\,\overline{f}(x)\ge\zeta\overline{f}(x\zeta)\}$.
\end{IEEEproof}
As a result, we can bound the log-rate and the $L^{p}$ error of an
asymptotically scale-invariant multi-resolution quantizer using the
following corollary and Proposition \ref{prop:lp_asymp}.
\begin{cor}
\label{cor:rate_error}For any asymptotically scale-invariant multi-resolution
quantizer $\{Q_{s}\}_{s>0}$ with finite log-rate, for any $s>0$,
$p>0$,
\begin{equation}
R_{0}(Q_{s})-R_{p+1}(Q_{s})\ge\frac{1}{p}\log_{2}\left(\frac{1-2^{-p}}{p}(\log_{2}e)^{p+1}\right).\label{eq:rate_error_tradeoff}
\end{equation}
\smallskip{}
\end{cor}
Recall that for the BBMRQ, by \eqref{eq:opt_approx_renyi}, $R_{0}(Q_{s})$
can be arbitrarily close to $\log_{2}\log_{2}e-\log_{2}s$, and $R_{p+1}(Q_{s})$
can be arbitrarily close to 
\[
-\frac{1}{p}\log_{2}\left(\frac{1-2^{-p}}{p}\log_{2}e\right)-\log_{2}s.
\]
Therefore the lower bound in \eqref{eq:rate_error_tradeoff} can be
approached. This shows that the tradeoff between log-rate and $L^{p}$
error of BBMRQ can be arbitrarily close to optimal.

Nevertheless, it is unknown whether the lower bound in \eqref{eq:rate_error_tradeoff}
can be attained exactly. It can be tracked in the proof of Corollary
\ref{cor:rate_error} that the equality in \eqref{eq:rate_error_tradeoff}
holds if and only if there exists $s>0$ such that $\overline{F}_{\{Q_{s}\}_{s}}(\gamma)=F_{2^{\mathrm{Unif}[-1,0]}}(\gamma s)$
for all $\gamma$, i.e., Theorem \ref{thm:opt_approx} can be attained
exactly. We conjecture that this is impossible.
\begin{conjecture}
There does not exist an asymptotically scale-invariant centered MRQ
$\{Q_{s}\}_{s>0}$ with $\overline{F}_{\{Q_{s}\}_{s}}=F_{2^{\mathrm{Unif}[-1,0]}}$.

\smallskip{}
\end{conjecture}
We now prove Corollary \ref{cor:rate_error}.
\begin{IEEEproof}
By Theorem \ref{thm:pos_neg}, for any pdf $\psi:\mathbb{R}_{>0}\to\mathbb{R}_{\ge0}$,
\begin{align*}
\int\bar{f}(y)\psi(y)\mathrm{d}y & \le\int_{0}^{\infty}\int_{0}^{\infty}x^{-1}\left(1+\mathbf{1}\{x>y\}\right)\bar{f}(x)\mathrm{d}x\cdot\psi(y)\mathrm{d}y\\
 & =\int_{0}^{\infty}x^{-1}\left(1+\int_{0}^{x}\psi(y)\mathrm{d}y\right)\bar{f}(x)\mathrm{d}x,
\end{align*}
\begin{equation}
\int_{0}^{\infty}\left(x^{-1}\left(1+\int_{0}^{x}\psi(y)\mathrm{d}y\right)-\psi(x)\right)\bar{f}(x)\mathrm{d}x\ge0.\label{eq:bd_intform}
\end{equation}
Substitute
\[
\psi(y)=\mathbf{1}\{1/2\le y\le1\}\frac{(1-2^{-p})\log_{2}y-y^{p}+1}{(1-2^{-p})\left(\frac{1-\log_{2}e}{2}\right)-\frac{1-2^{-(p+1)}}{p+1}+1/2}.
\]
For $1/2\le x\le1$,
\begin{align*}
 & x^{-1}\left(1+\int_{0}^{x}\psi(y)\mathrm{d}y\right)-\psi(x)\\
 & =x^{-1}+\frac{x^{-1}\int_{1/2}^{x}\left((1-2^{-p})\log_{2}y-y^{p}+1\right)\mathrm{d}y-\left((1-2^{-p})\log_{2}x-x^{p}+1\right)}{(1-2^{-p})\left(\frac{1-\log_{2}e}{2}\right)-\frac{1-2^{-(p+1)}}{p+1}+1/2}\\
 & =x^{-1}+\frac{x^{-1}\left((1-2^{-p})\left(-x\log_{2}e+x\log_{2}x+(\log_{2}e)/2+1/2\right)-\frac{x^{p+1}-2^{-(p+1)}}{p+1}+x-1/2\right)-\left((1-2^{-p})\log_{2}x-x^{p}+1\right)}{(1-2^{-p})\left(\frac{1-\log_{2}e}{2}\right)-\frac{1-2^{-(p+1)}}{p+1}+1/2}\\
 & =x^{-1}+\frac{(1-2^{-p})\left(-\log_{2}e+\log_{2}x+(\log_{2}e)x^{-1}/2+x^{-1}/2\right)-\frac{x^{p}-2^{-(p+1)}x^{-1}}{p+1}+1-x^{-1}/2-(1-2^{-p})\log_{2}x+x^{p}-1}{(1-2^{-p})\left(\frac{1-\log_{2}e}{2}\right)-\frac{1-2^{-(p+1)}}{p+1}+1/2}\\
 & =x^{-1}+\frac{(1-2^{-p})\left(-\log_{2}e+(\log_{2}e)x^{-1}/2+x^{-1}/2\right)-\frac{x^{p}-2^{-(p+1)}x^{-1}}{p+1}-x^{-1}/2+x^{p}}{(1-2^{-p})\left(\frac{1-\log_{2}e}{2}\right)-\frac{1-2^{-(p+1)}}{p+1}+1/2}\\
 & =\left(1+\frac{(1-2^{-p})\left((\log_{2}e)/2+1/2\right)+\frac{2^{-(p+1)}}{p+1}-1/2}{(1-2^{-p})\left(\frac{1-\log_{2}e}{2}\right)-\frac{1-2^{-(p+1)}}{p+1}+1/2}\right)x^{-1}+\frac{(1-2^{-p})\left(-\log_{2}e\right)-\frac{x^{p}}{p+1}+x^{p}}{(1-2^{-p})\left(\frac{1-\log_{2}e}{2}\right)-\frac{1-2^{-(p+1)}}{p+1}+1/2}\\
 & =\left(\frac{\frac{p}{p+1}(1-2^{-p})}{(1-2^{-p})\left(\frac{1-\log_{2}e}{2}\right)-\frac{1-2^{-(p+1)}}{p+1}+1/2}\right)x^{-1}+\frac{(1-2^{-p})\left(-\log_{2}e\right)+x^{p}\frac{p}{p+1}}{(1-2^{-p})\left(\frac{1-\log_{2}e}{2}\right)-\frac{1-2^{-(p+1)}}{p+1}+1/2}\\
 & =\frac{\frac{p}{p+1}\left((1-2^{-p})x^{-1}+x^{p}\right)-(1-2^{-p})\log_{2}e}{(1-2^{-p})\left(\frac{1-\log_{2}e}{2}\right)-\frac{1-2^{-(p+1)}}{p+1}+1/2}\\
 & =\frac{\frac{p}{p+1}\left((1-2^{-p})x^{-1}+x^{p}\right)-(1-2^{-p})\log_{2}e}{(1-2^{-p})\left(\frac{1-\log_{2}e}{2}\right)+\frac{p/2-(1/2)(1-2^{-p})}{p+1}}\\
 & =2\cdot\frac{\frac{p}{p+1}\left((1-2^{-p})x^{-1}+x^{p}\right)-(1-2^{-p})\log_{2}e}{\frac{p}{p+1}(2-2^{-p})-(1-2^{-p})\log_{2}e}
\end{align*}
For $x\ge1$, we have
\[
p\ge(1-2^{-p})\log_{2}e,
\]
\[
\frac{p+1}{p}\ge\frac{p+1}{p^{2}}(1-2^{-p})\log_{2}e,
\]
\begin{align*}
x^{p+1}\ge1 & \ge\frac{p+1}{p^{2}}\left((1-2^{-p})\log_{2}e-\frac{p}{p+1}\right),
\end{align*}
\[
\frac{p^{2}}{p+1}x^{p-1}\ge\left((1-2^{-p})\log_{2}e-\frac{p}{p+1}\right)x^{-2}.
\]
Integrating both sides from $1$ to $x$,
\[
\frac{p}{p+1}(x^{p}-1)\ge\left((1-2^{-p})\log_{2}e-\frac{p}{p+1}\right)(1-x^{-1}),
\]
\[
\left((1-2^{-p})\log_{2}e-\frac{p}{p+1}\right)x^{-1}+\frac{p}{p+1}x^{p}-(1-2^{-p})\log_{2}e\ge0.
\]
Therefore,
\begin{align*}
 & 2\cdot\frac{\frac{p}{p+1}\left((1-2^{-p})x^{-1}+x^{p}\right)-(1-2^{-p})\log_{2}e}{\frac{p}{p+1}(2-2^{-p})-(1-2^{-p})\log_{2}e}-\left(x^{-1}\left(1+\int_{0}^{x}\psi(y)\mathrm{d}y\right)-\psi(x)\right)\\
 & =2\cdot\frac{\frac{p}{p+1}\left((1-2^{-p})x^{-1}+x^{p}\right)-(1-2^{-p})\log_{2}e}{\frac{p}{p+1}(2-2^{-p})-(1-2^{-p})\log_{2}e}-2x^{-1}\\
 & =2\cdot\frac{\left((1-2^{-p})\log_{2}e-\frac{p}{p+1}\right)x^{-1}+\frac{p}{p+1}x^{p}-(1-2^{-p})\log_{2}e}{\frac{p}{p+1}(2-2^{-p})-(1-2^{-p})\log_{2}e}\\
 & \ge0.
\end{align*}
For $0<x\le1/2$, by the concavity of $y\mapsto y\log_{2}(1/y)$,
\[
p2^{-p}=2^{-p}\log_{2}2^{p}\le(1-2^{-p})\log_{2}e,
\]
\[
\frac{p+1}{p}2^{-(p+1)}\le\frac{p+1}{2p^{2}}(1-2^{-p})\log_{2}e,
\]
\[
x^{p+1}\le\frac{p+1}{2p^{2}}(1-2^{-p})\log_{2}e-\frac{1}{2p}2^{-p},
\]
\[
x^{p+1}\le\frac{p+1}{2p^{2}}\left((1-2^{-p})\log_{2}e-\frac{p}{p+1}2^{-p}\right),
\]
\[
\frac{2p^{2}}{p+1}x^{p-1}\le\left((1-2^{-p})\log_{2}e-\frac{p}{p+1}2^{-p}\right)x^{-2},
\]
Integrating both sides from $x$ to $1/2$,
\[
\frac{2p}{p+1}(2^{-p}-x^{p})\le\left((1-2^{-p})\log_{2}e-\frac{p}{p+1}2^{-p}\right)(x^{-1}-2),
\]
\[
\left((1-2^{-p})\log_{2}e-\frac{p}{p+1}2^{-p}\right)x^{-1}+\frac{2p}{p+1}x^{p}-2(1-2^{-p})\log_{2}e\ge0.
\]
Therefore,
\begin{align*}
 & 2\cdot\frac{\frac{p}{p+1}\left((1-2^{-p})x^{-1}+x^{p}\right)-(1-2^{-p})\log_{2}e}{\frac{p}{p+1}(2-2^{-p})-(1-2^{-p})\log_{2}e}-\left(x^{-1}\left(1+\int_{0}^{x}\psi(y)\mathrm{d}y\right)-\psi(x)\right)\\
 & =2\cdot\frac{\frac{p}{p+1}\left((1-2^{-p})x^{-1}+x^{p}\right)-(1-2^{-p})\log_{2}e}{\frac{p}{p+1}(2-2^{-p})-(1-2^{-p})\log_{2}e}-x^{-1}\\
 & =\frac{\left((1-2^{-p})\log_{2}e-\frac{p}{p+1}2^{-p}\right)x^{-1}+\frac{2p}{p+1}x^{p}-2(1-2^{-p})\log_{2}e}{\frac{p}{p+1}(2-2^{-p})-(1-2^{-p})\log_{2}e}\\
 & \ge0.
\end{align*}
Hence, for any $x>0$,
\[
x^{-1}\left(1+\int_{0}^{x}\psi(y)\mathrm{d}y\right)-\psi(x)\le\frac{\frac{p}{p+1}\left((1-2^{-p})x^{-1}+x^{p}\right)-(1-2^{-p})\log_{2}e}{\frac{p}{p+1}(2-2^{-p})-(1-2^{-p})\log_{2}e}.
\]
By \eqref{eq:bd_intform},
\[
\frac{p(1-2^{-p})}{p+1}\int_{0}^{\infty}x^{-1}\bar{f}(x)\mathrm{d}x+\frac{p}{p+1}\int_{0}^{\infty}x^{p}\bar{f}(x)\mathrm{d}x\ge(1-2^{-p})\log_{2}e,
\]
\[
(1-2^{-p})\int_{0}^{\infty}x^{-1}\bar{f}(x)\mathrm{d}x+\int_{0}^{\infty}x^{p}\bar{f}(x)\mathrm{d}x\ge\frac{p+1}{p}(1-2^{-p})\log_{2}e.
\]
Therefore,
\[
(1-2^{-p})2^{R_{0}(Q_{1})}+2^{-pR_{p+1}(Q_{1})}\ge\frac{p+1}{p}(1-2^{-p})\log_{2}e.
\]
Note that the above also holds when $Q_{1}$ is replaced by $Q_{s}$.
Since $R_{\eta}(Q_{s})=R_{\eta}(Q_{1})-\log_{2}s$, we have, for any
$s>0$,
\[
(1-2^{-p})2^{R_{0}(Q_{1})-\log_{2}s}+2^{-p(R_{p+1}(Q_{1})-\log_{2}s)}\ge\frac{p+1}{p}(1-2^{-p})\log_{2}e,
\]
\[
(1-2^{-p})s^{-1}2^{R_{0}(Q_{1})}+s^{p}2^{-pR_{p+1}(Q_{1})}\ge\frac{p+1}{p}(1-2^{-p})\log_{2}e.
\]
Substituting
\[
s=\left(\frac{1}{p}(1-2^{-p})2^{R_{0}(Q_{1})+pR_{p+1}(Q_{1})}\right)^{\frac{1}{p+1}},
\]
we have
\begin{align*}
 & \frac{p+1}{p}(1-2^{-p})\log_{2}e\\
 & \le(1-2^{-p})s^{-1}2^{R_{0}(Q_{1})}+s^{p}2^{-pR_{p+1}(Q_{1})}\\
 & =(1-2^{-p})\left(\frac{1}{p}(1-2^{-p})2^{R_{0}(Q_{1})+pR_{p+1}(Q_{1})}\right)^{-\frac{1}{p+1}}2^{R_{0}(Q_{1})}+\left(\frac{1}{p}(1-2^{-p})2^{R_{0}(Q_{1})+pR_{p+1}(Q_{1})}\right)^{\frac{p}{p+1}}2^{-pR_{p+1}(Q_{1})}\\
 & =(1-2^{-p})^{\frac{p}{p+1}}p^{\frac{1}{p+1}}2^{\frac{p}{p+1}R_{0}(Q_{1})-\frac{p}{p+1}R_{p+1}(Q_{1})}+(1-2^{-p})^{\frac{p}{p+1}}p^{-\frac{p}{p+1}}2^{\frac{p}{p+1}R_{0}(Q_{1})-\frac{p}{p+1}R_{p+1}(Q_{1})}\\
 & =\left((1-2^{-p})2^{R_{0}(Q_{1})-R_{p+1}(Q_{1})}\right)^{\frac{p}{p+1}}\left(p^{\frac{1}{p+1}}+p^{-\frac{p}{p+1}}\right).
\end{align*}
Hence,
\begin{align*}
\left((1-2^{-p})2^{R_{0}(Q_{1})-R_{p+1}(Q_{1})}\right)^{\frac{p}{p+1}} & \ge\frac{(p+1)(1-2^{-p})\log_{2}e}{p^{1+\frac{1}{p+1}}+p^{\frac{1}{p+1}}}\\
 & =p^{-\frac{1}{p+1}}(1-2^{-p})\log_{2}e,
\end{align*}
\[
(1-2^{-p})2^{R_{0}(Q_{1})-R_{p+1}(Q_{1})}\ge p^{-\frac{1}{p}}\left((1-2^{-p})\log_{2}e\right)^{\frac{p+1}{p}},
\]
\[
2^{R_{0}(Q_{1})-R_{p+1}(Q_{1})}\ge p^{-\frac{1}{p}}(1-2^{-p})^{\frac{1}{p}}(\log_{2}e)^{\frac{p+1}{p}},
\]
\begin{align*}
R_{0}(Q_{1})-R_{p+1}(Q_{1}) & \ge\log_{2}\left(p^{-\frac{1}{p}}(1-2^{-p})^{\frac{1}{p}}(\log_{2}e)^{\frac{p+1}{p}}\right)\\
 & =\frac{1}{p}\log_{2}\left(\frac{1-2^{-p}}{p}(\log_{2}e)^{p+1}\right).
\end{align*}
\end{IEEEproof}

\medskip{}

\[
\]

 \bibliographystyle{IEEEtran}
\bibliography{ref}

\begin{thebibliography}{10}
\providecommand{\url}[1]{#1}
\csname url@samestyle\endcsname
\providecommand{\newblock}{\relax}
\providecommand{\bibinfo}[2]{#2}
\providecommand{\BIBentrySTDinterwordspacing}{\spaceskip=0pt\relax}
\providecommand{\BIBentryALTinterwordstretchfactor}{4}
\providecommand{\BIBentryALTinterwordspacing}{\spaceskip=\fontdimen2\font plus
\BIBentryALTinterwordstretchfactor\fontdimen3\font minus
  \fontdimen4\font\relax}
\providecommand{\BIBforeignlanguage}[2]{{%
\expandafter\ifx\csname l@#1\endcsname\relax
\typeout{** WARNING: IEEEtran.bst: No hyphenation pattern has been}%
\typeout{** loaded for the language `#1'. Using the pattern for}%
\typeout{** the default language instead.}%
\else
\language=\csname l@#1\endcsname
\fi
#2}}
\providecommand{\BIBdecl}{\relax}
\BIBdecl

\bibitem{koshelev1980hierarchical}
V.~Koshelev, ``Hierarchical coding of discrete sources,'' \emph{Problemy
  peredachi informatsii}, vol.~16, no.~3, pp. 31--49, 1980.

\bibitem{equitz1991successive}
W.~H. Equitz and T.~M. Cover, ``Successive refinement of information,''
  \emph{IEEE Transactions on Information Theory}, vol.~37, no.~2, pp. 269--275,
  1991.

\bibitem{rimoldi1994successive}
B.~Rimoldi, ``Successive refinement of information: Characterization of the
  achievable rates,'' \emph{IEEE Transactions on Information Theory}, vol.~40,
  no.~1, pp. 253--259, 1994.

\bibitem{brunk1996fixedrate}
H.~{Brunk} and N.~{Farvardin}, ``Fixed-rate successively refinable scalar
  quantizers,'' in \emph{Proceedings of Data Compression Conference - DCC '96},
  March 1996, pp. 250--259.

\bibitem{brunk1997entropy}
H.~{Jafarkhani}, H.~{Brunk}, and N.~{Farvardin}, ``Entropy-constrained
  successively refinable scalar quantization,'' in \emph{Proceedings DCC '97.
  Data Compression Conference}, March 1997, pp. 337--346.

\bibitem{effros1998practical}
M.~Effros, ``Practical multi-resolution source coding: {TSVQ} revisited,'' in
  \emph{Proceedings DCC'98 Data Compression Conference (Cat. No.
  98TB100225)}.\hskip 1em plus 0.5em minus 0.4em\relax IEEE, 1998, pp. 53--62.

\bibitem{wu2002multires}
{Xiaolin Wu} and S.~{Dumitrescu}, ``On optimal multi-resolution scalar
  quantization,'' in \emph{Proceedings DCC 2002. Data Compression Conference},
  April 2002, pp. 322--331.

\bibitem{dumitrescu2004algorithms}
S.~Dumitrescu and X.~Wu, ``Algorithms for optimal multi-resolution
  quantization,'' \emph{Journal of Algorithms}, vol.~50, no.~1, pp. 1 -- 22,
  2004.

\bibitem{effros2004multires}
M.~{Effros} and D.~{Dugatkin}, ``Multiresolution vector quantization,''
  \emph{IEEE Transactions on Information Theory}, vol.~50, no.~12, pp.
  3130--3145, Dec 2004.

\bibitem{wolf1980source}
J.~K. Wolf, A.~D. Wyner, and J.~Ziv, ``Source coding for multiple
  descriptions,'' \emph{The Bell System Technical Journal}, vol.~59, no.~8, pp.
  1417--1426, 1980.

\bibitem{ozarow1980source}
L.~Ozarow, ``On a source-coding problem with two channels and three
  receivers,'' \emph{Bell System Technical Journal}, vol.~59, no.~10, pp.
  1909--1921, 1980.

\bibitem{gamal1982achievable}
A.~{El Gamal} and T.~Cover, ``Achievable rates for multiple descriptions,''
  \emph{IEEE Transactions on Information Theory}, vol.~28, no.~6, pp. 851--857,
  1982.

\bibitem{yeung1995multilevel}
R.~W. Yeung, ``Multilevel diversity coding with distortion,'' \emph{IEEE
  Transactions on Information Theory}, vol.~41, no.~2, pp. 412--422, 1995.

\bibitem{vaishampayan1993design}
V.~A. Vaishampayan, ``Design of multiple description scalar quantizers,''
  \emph{IEEE Transactions on Information Theory}, vol.~39, no.~3, pp. 821--834,
  1993.

\bibitem{lloyd1982least}
S.~Lloyd, ``Least squares quantization in {PCM},'' \emph{IEEE transactions on
  information theory}, vol.~28, no.~2, pp. 129--137, 1982.

\bibitem{max1960quantizing}
J.~Max, ``Quantizing for minimum distortion,'' \emph{IRE Transactions on
  Information Theory}, vol.~6, no.~1, pp. 7--12, 1960.

\bibitem{ross1995stochastic}
S.~M. Ross, \emph{Stochastic processes}, 2nd~ed.\hskip 1em plus 0.5em minus
  0.4em\relax New York: Wiley, 1995.

\end{thebibliography}

\end{document}